# Lightest to Heavy UHECR nuclei in a Local Universe

*Daniele* Fargion[1,2,*]

[1]Physics Department, Rome University 1, Sapienza, Italy;
[2]Osservatorio Astronomico di Capodimonte, INAF, Italy,

**Abstract.** UHECR are evalueted in the frame role of different nuclei composition . Most of the past and present models are considering proton or iron as their main currier. Some attention has been paid to the role of the UHECR light nuclei in recent years. We update here the lightest nuclei UHECR model, able to explain the nearest AGN or Star Burst sources with the few observed Hot Spot clustering in AUGER and TA array data. Any additional components of the heaviest nuclei with the highest energy, more bent and smeared, may also fit recent AUGER and TA homogeneous records at those energy edges.

## 1 Introduction

Cosmic Rays, CR, below ten or hundred PeV, ($10^{16} - 10^{17}$ eV) energy, are expected to be mostly galactic particles, spiraling and contained by galactic fields inside our Milky Way [3]. At higher nuclei energy , ($E_{CR} >> 10^{18}$ eV), such ultrahigh energy cosmic ray (UHECR), are expected by most authors, to be originated by far extra galactic sources. However, because of the expected photo-nucleon opacity, due to photo-pion production, the most energetic UHECR , above ($E_{CR} > 4 \cdot 10^{19}$ eV) energy, were suggested since 1966 to suffer of a severe spectra distribution fall down, the so called GZK cut off [1], [2]. Such a cut-off had been revealed much later, on 2007 by Hires array [5] and soon later by AUGER [6]. Because of it, most authors favored a proton UHECR nature. Such nucleon UHECR constrained by a GZK cut off, could reach us only within nearly (1%) of the Universe radius, Or nearly within a 40 Mpc distance. Such proton UHECR models , with the first two dozen of events in AUGER data, on 2007, [6]. raised a new question : how , the expected UHECR events originated , by Virgo Cluster , (thousands of galaxies sources at near 25 Mpc), within GZK cut off , be unobserved ? Such a puzzling absence promptly inspired the Lightest nuclei UHECR model [4]. Indeed lightest UHECR nuclei are also suffering an earlier and more restrictive GZK cut off, the one due to a photo-nuclear destruction of these lightest He-like nuclei, via GDR , (the Giant Dipole Resonance). An opacity both in the spectra and in the survival distance. A corresponding limiting distance of only (0.1% of the cosmic radius) or within a 4 Mpc radius. Such a filter , for lightest nuclei, could hide the Virgo Cluster signals, solving its puzzling absence for proton nucleon. These lightest nuclei confinement distance is not just ten times smaller than the nucleon GZK one: it imply a volume thousand times smaller , with only a few, known sources : the Local Sheet Galaxies. The lightest (to light) composition key role, generally ignored in last decade, had been widely agreed only recently (2018) [10]. The slanth depth shape of UHECR had shown such lightest to heavy signature above ten EeV energy. The Lightest nuclei model fit a very local UHECR Universe, that is well inside our Local Sheet Galaxy group, containing a dozen of nearby galaxies. All of them are somehow correlated with the recent UHECR clustering, mostly toward Cen A, M82, NGC 253, and Cas A, Maffei galaxies and M31, our nearest Andromeda companion [7]. The Lightest UHECR nuclei as *D*, *He*, *Li*, *Be*, because their small charges, allow to the UHECR to mantain a partial directionality , offering a first UHECR smeared astronomy. Heavier light (C,O,N) or heaviest (Fe, Ni) nuclei, are much charged and smeared. They are not much constrained in near local distances and cannot keep much memory of their origin. Therefore heaviest nuclei like Fe, Ni, (more deflected and smeared) might explain the most recent AUGER and TA homogeneous maps at their highest energy ($> 8 \cdot 10^{19}$ eV) . Finally , the detected AUGER dipole anisotropy around lower ($10^{19}$ eV) energy, could also be explained by the lightest nuclei (partial polluted by galactic sources as LMC and Vela) ruled by the main mix contribute by Cen A, and NGC 253 AGN. The Galactic microquasars, as SS433 might also play a very interesting key roles . Indeed its possible nature, as a tens *PeV* neutron jet [8], [9], could also lead , at its maximal energy, to an additional UHECR contribute , correlated with a rare quadruplet clustering of events, observed by AUGER and TA. [7]. We remind finally the exotic solution , based on UHE neutrino interactions with cosmic relic neutrinos, nearly at rest, having a light eV mass, explaining rare puzzling UHECR events. That model, often recalled as Z-resonance or Z-burst model, is based on highest, ZeV ($E_\nu >> 10^{21}$ eV) energetic neutrinos hitting on relic ones at rest in dark halos [11] , avoiding any GZK cut-off. Their ultra-relativistic Z-boson decay produce the observed UHECR nucleons (and anti nucleons) correlating with most far cosmic sources.

[*]e-mail: daniele.fargion@fondazione.uniroma1.it

## 2 Conclusions

The UHECR , by their hardness and directionality, are offering a first windows to the source origination and a rare view to highest energy astronomy. The observed lightest to heavy nuclei presence in UHECR , their increasing masses and charges with growing UHECR energy , makes only lightest nuclei less bent with some clustering ability . This occur only around a few tens of EeV energy. These lightest nuclei , D,He,Li,Be , their fragility, defines a very narrow UHECR Universe, located within few Mpc, leading to very few local sources. We believe, that such a desired rare astronomy, by AUGER , TA widest arrays, is now arising [7].

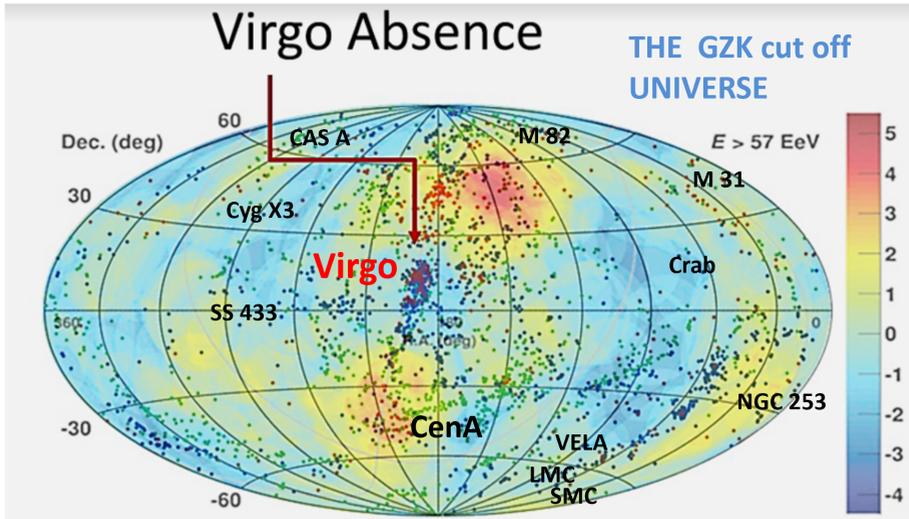

**Figure 1.** The earliest Virgo cluster expected role in celestial coordinate. The central Virgo cluster dominance , well within the GZK photo-pion opacity, is absent in any UHECR clustering records . The North UHECR Hot spot clustering toward M82, and one , in South sky, toward CenA, are both the nearest AGN found in the earliest UHECR data . The nearer red, (or blue) and the far green dots , are signals of thousands of Virgo galaxies, all within 40 Mpc GZK volume, all reachable to us for UHECR nucleon. But not for more constrained He-like nuclei. [7].

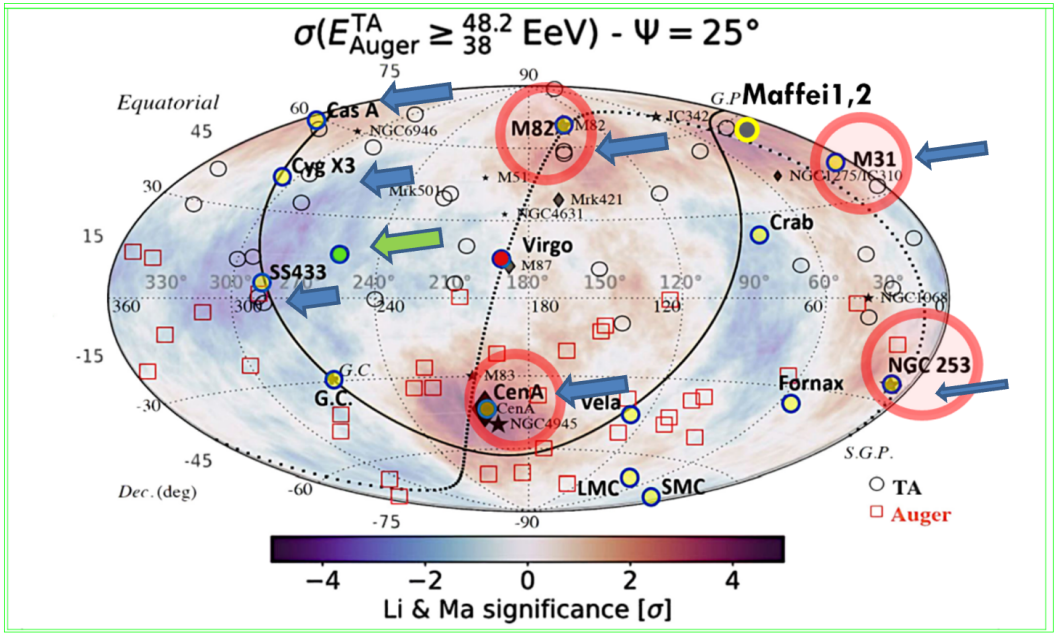

**Figure 2.** As in previous figure. the updated UHECR clustering with the recognized nearest galaxies within the Local Sheet. The blue arrow point to the nearby galaxies as well as SS433 and Cygnus X3 microquasars, while the green arrow signal the last unexplained UHECR "Amaterasu" event. [7].